\documentclass{jpsj-suppl}
\usepackage{txfonts} %Please comment out this line unless the txfonts package is availabe in your LaTeX system.

\usepackage{bm}

\usepackage{color}

\newcommand{\ET}{$\alpha$-(BEDT-TTF)$_2$I$_3$}
\newcommand{\up}{\uparrow}
\newcommand{\dn}{\downarrow}
\newcommand{\Va}{V_\mathrm{a}}
\newcommand{\Vb}{V_\mathrm{b}}

\title{Edge States in Molecular Solid \ET: Effects of Electron Correlations}

\author{Yukiko \textsc{Omori}$^{1,2}$, Genki \textsc{Matsuno}$^{1}$ and Akito \textsc{Kobayashi}$^1$}

\inst{$^{1}$Department of Physics, Nagoya University, Nagoya 464-8602, Japan \\
$^{2}$Institute for Advanced Research, Nagoya University, Nagoya 464-8602, Japan}

\email{omori@s.phys.nagoya-u.ac.jp}

\recdate{July 18, 2013}

\abst{
We examine the edge states of the Dirac electrons in the molecular material \ET
\ with electron-electron interactions.
Based on the analysis of the extended Hubbard model with the Hartree-Fock approximation,
we show that the charge-ordered phase has the gapless edge states
only in the vicinity of the phase boundary between the zero-gap state and the charge-ordered phase.
We also show a peculiar flux phase caused by the long-range Coulomb interaction
with the same mechanism as topological Mott insulator phase.
}

\kword{Organic conductor, Dirac electrons, edge state, charge ordering}

\begin{document}
\maketitle

\section{Introduction}

The Dirac electrons in condensed matters have attracted attention in basic science and device application.
It is found in several materials, e.g., graphene and bismuth alloy.
The quasi-two-dimensional organic conductor \ET \ is the first bulk crystal 
where the massless Dirac electrons are realized:
It exhibits the zero-gap state (ZGS) under pressure 
\cite{Tajima_JPSJ2006,Katayama_JPSJ2006}.
Compared with other materials,
it is worthy of notice that this system has strong long-range Coulomb interaction due to the weak screening.
Actually, the nearest-neighbor Coulomb repulsion in this system causes a metal-insulator transition
at $T_\mathrm{MI}=$135K at ambient pressure 
\cite{Bender_MCLC1984}.
The insulating phase is the stripe charge ordered (CO) phase confirmed by NMR experiment 
\cite{Takahashi_SM2003}
as theoretically proposed 
\cite{Kino_JPSJ1995,Seo_JPSJ2000}.
Effects of electron correlation is considerable when examining the physical nature of \ET.

One of the characteristic features of Dirac electrons in materials appears in the edge state.
In \ET, it is shown that the edge states exist even under the artificial breaking of the inversion symmetry
\cite{Hasegawa_JPSJ2011},
where the interaction terms, however, has not been taken into account.
Because the gapless edge state can yield some contributions to the transport characteristics,
it is important to study the edge states in the realistic ordered phase with ineteractions.
For example, in the vicinity of the phase boundary between ZGS and CO,
the exceptional behavior of the Seebeck coefficient is observed
\cite{Tajima_private2012}.
It implies existence of unconventional conduction channels.
Moreover, in another aspect,
it is intriguing to examine the possible topological nature in the \ET.
In topological insulators the edge states are topologically protected and
robust to the non-magnetic perturbations.
The major mechanism of the topological insulators originates in spin-orbit interaction \cite{Kane_PRL2005}.
However, it is shown that the Fock terms of inter-site repulsions also can cause the topological insulator phase,
so called ``topological Mott insulator''\cite{Raghu_PRL2008}
(Note that there is another phase referred as the same \cite{Pesin_NP2010}).
Although the most possible candidate in organic conductors having this interaction-driven topological phase is \ET,
but it has not been studied up to the present.

In this paper,
we investigate the effect of the electron-electron correlation on the edge state in \ET \ theoretically.
In section II, we introduce the model and scheme.
In section III, we show the numerical results.
Section IV is devoted to summary and discussion.

%-----------------------------------------------------------------------------------------------------------------
\section{Model and Scheme}
%-----------------------------------------------------------------------------------------------------------------

\subsection{Bulk System}

\begin{figure}[b]
 \centering
 \includegraphics[height=2in]{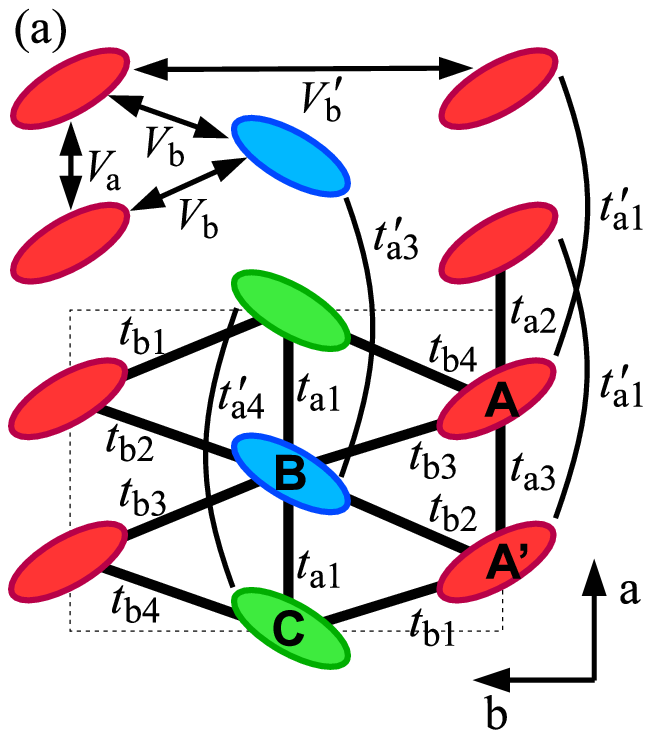}
 \hspace*{10mm}
 \includegraphics[height=2in]{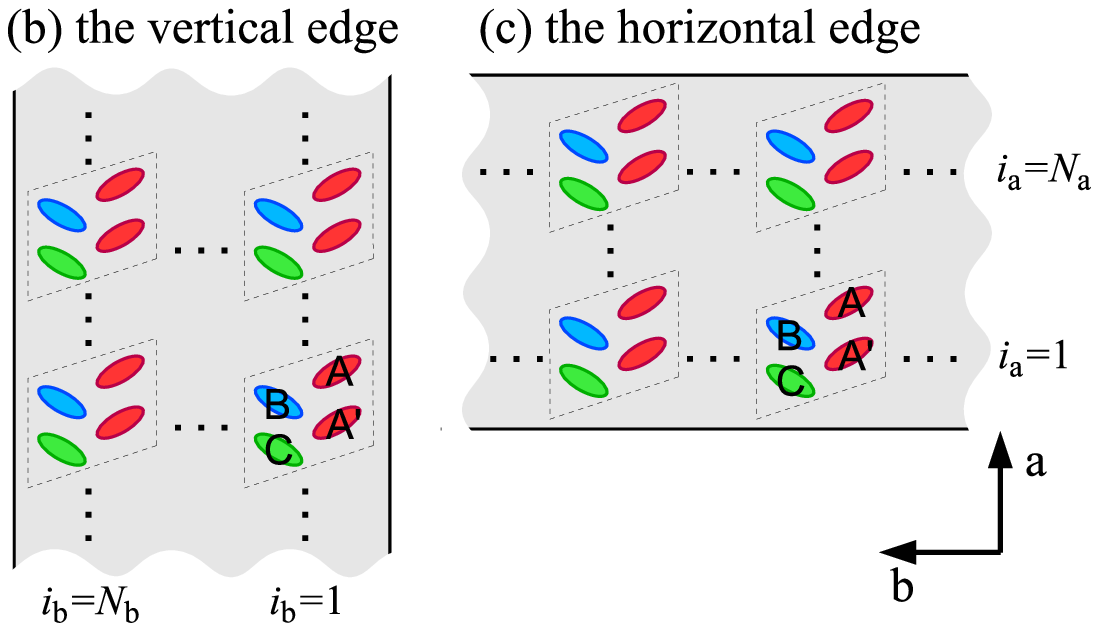}
 \caption{
 (a) Lattice structure of \ET.
 The unit cell consist of four BEDT-TTF molecules, A, A', B and C,
 and ten inequivalent transfer integrals, $t_{\mathrm{a}i}$, $t_{\mathrm{b}i}$, and $t'_{\mathrm{a}i}$
 \cite{Kino_JPSJ2006}.
 (b) The semi-infinite system with the vertical edge, and
 (c) the horizontal edge.
 }
 \label{fig:lattice}
\end{figure}

The microscopic model for \ET \ is given by the two-dimensional extended Hubbard model 
on a strongly anisotropic triangular lattice described in Fig.~\ref{fig:lattice}(a),
%====================
\begin{equation}
 H = \sum_{(i\alpha,j\beta), s} t_{i,\alpha;j,\beta} \ c_{i,\alpha,s}^\dagger c_{j,\beta,s}
  + \sum_{i\alpha} U \ n_{i,\alpha,\up} n_{i,\alpha,\dn}
  + \sum_{(i\alpha,j\beta)} V_{i,\alpha;j,\beta} \ n_{i,\alpha} n_{j,\beta},
  \label{eq:Hami}
\end{equation}
%====================
where $c_{i,\alpha,s}^\dagger$ is the electron creation operator with spin $s$ ($=\up,\dn$)
on the $\alpha$ ($=$A,A',B, and C) molecule in the $i$-th unit cell.
We define the number operator $n_{i,\alpha,s}=c_{i,\alpha,s}^\dagger c_{i,\alpha,s}$
and $n_{i,\alpha}=n_{i,\alpha,\up} + n_{i,\alpha,\dn}$.
$U$ is the onsite repulsion, and in this paper we fix $U=0.4$ eV.
The quantities $t_{i,\alpha;j,\beta}$ and $V_{i,\alpha;j,\beta}$ are
the transfer energy and the inter-site Coulomb repulsion between $(i,\alpha)$ and $(j,\beta)$ sites, respectively.
For the transfer energies,
we adopt the numerically estimated values with the \textit{ab initio} calculation 
at $T=8$K under  ambient pressure \cite{Kino_JPSJ2006}:
$t_\mathrm{a1}  = -0.0267$,
$t_\mathrm{a2}  = -0.0511$,
$t_\mathrm{a3}  =  0.0323$,
$t_\mathrm{b1}  =  0.1241$,
$t_\mathrm{b2}  =  0.1296$,
$t_\mathrm{b3}  =  0.0513$,
$t_\mathrm{b4}  =  0.0152$,
$t_\mathrm{a1}' =  0.0119$,
$t_\mathrm{a3}' =  0.0046$, and
$t_\mathrm{a4}' =  0.0060$ with eV as units
(see Fig.~\ref{fig:lattice}(a)).
As for the inter-site repulsion $V_{i,\alpha;j,\beta}$, 
we consider the three kinds of the values, $\Va$, $\Vb$, and $\Vb'$,
as shown in Fig.~\ref{fig:lattice}(a).
%$\Va$ and $\Vb$ are the nearst-neighbor repulsions along the a and b directions, respectively.
In this paper we set $\Vb=0.7\Va$ from a rough estimation
of the distance ratio between the two molecules.
The next-nearest repulsion along the b axis, $\Vb'$, is expected to be small in the realistic system.
However, this long-range interaction plays indispensable role 
on the appearance of a peculiar flux phase related to the topological Mott insulator, as shown in section 3.2.
In the following calculation, we fix the temperature at $k_\mathrm{B}T= 0.0001$ eV,
and consider the $3/4$-filled system.

%====================
%====================

We treat this model (\ref{eq:Hami}) with the Hartree-Fock approximation
for the bulk system, i.e., under the periodic boundary condition.
The interaction terms are decoupled to
%====================
\begin{eqnarray}
 n_{i,\alpha,s} n_{j,\beta,s'}
  & \sim &
  \big< n_{i,\alpha,s} \big> n_{j,\beta,s'} + n_{i,\alpha,s} \big< n_{j,\beta,s'} \big>
  -\big< n_{i,\alpha,s} \big> \big< n_{j,\beta,s'} \big>
\nonumber \\ &&
  - \big< c_{i,\alpha,s}^\dagger c_{j,\beta,s'} \big> c_{j,\beta,s'}^\dagger c_{i,\alpha,s}
  - c_{i,\alpha,s}^\dagger c_{j,\beta,s'} \big< c_{j,\beta,s'}^\dagger c_{i,\alpha,s} \big>
  + \big< c_{i,\alpha,s}^\dagger c_{j,\beta,s'} \big> \big< c_{j,\beta,s'}^\dagger c_{i,\alpha,s} \big>.
  \label{eq:HF}
\end{eqnarray}
%====================
We calculate the mean fields $\big< c_{i,\alpha,s}^\dagger c_{j,\beta,s'} \big>$ itinerantly
preserving the translational symmetry.
We can summarize the resultant mean-field Hamiltonian in the following one-body one as
%====================
\begin{equation}
 H_\mathrm{MF} =
  \sum_{i\alpha} \epsilon_{\alpha} n_{i,\alpha}
  - 2U \sum_{i\alpha} \left< {\bm S}_{i,\alpha} \right> \cdot {\bm S}_{i,\alpha}
  + \sum_{(i\alpha,j\beta), s} {\tilde t}^\mathrm{\ c}_{i,\alpha;j,\beta} \ c_{i,\alpha,s}^\dagger c_{j,\beta,s}
  + \sum_{(i\alpha,j\beta)} {\tilde {\bm t}}^\mathrm{\ s}_{i,\alpha;j,\beta} \cdot
  \sum_{ss'} c_{i,\alpha,s}^\dagger {\bm \sigma}_{ss'} c_{j,\beta,s'} ,
  \label{eq:HMF}
\end{equation}
%====================
where ${\bm \sigma}$ is the Pauli matrix and 
${\bm S}_{i,\alpha} = \frac{1}{2} \sum_{ss'} c_{i,\alpha,s}^\dagger {\bm \sigma}_{ss'} c_{i,\alpha,s'} $.
The spin-independent effective onsite potential $\epsilon_\alpha$ comes from the Hartree terms of Eq.~(\ref{eq:HF}).
The quantity ${\tilde t}^\mathrm{\ c}_{i,\alpha;j,\beta}$ (${\bm {\tilde t}}^\mathrm{\ s}_{i,\alpha;j,\beta}$)
is the spin-independent (-dependent) transfer integral consisting of the intrinsic transfer $t_{i,\alpha;j,\beta}$
and the Fock terms of the inter-site repulsions $\Va$, $\Vb$, and $\Vb'$.
Note that the renormalized transfer integrals 
${\tilde t}^\mathrm{\ c}_{i,\alpha;j,\beta}$ and ${\bm {\tilde t}}^\mathrm{\ s}_{i,\alpha;j,\beta}$
can become the complex number when the resultant Fock terms has the imaginary part.

\subsection{Edge States}

In order to study the edge states in ordered phases caused by interactions,
we consider the semi-infinite systems with the vertical or horizontal edges 
as shown in Fig.~\ref{fig:lattice} (b) and (c), respectively.
We impose the periodic boundary condition on the system only along the a (b) direction
for the vertical (horizontal) edges.
To this semi-infinite systems
we adopt the resultant mean-field Hamiltonian (\ref{eq:HMF}) as the effective tight-binding Hamiltonian.
After the diagonalization, we can obtain the energy spectrum and its eigen vector 
$\psi_{\nu,i_b,\alpha}(k_a)$ ($\psi_{\nu,i_a,\alpha}(k_b)$) for the vertical (horizontal) edges,
where we define the band index $\nu$, the unit-cell index $i_b$ ($i_a$) along b (a) axis,
as shown in Fig.~\ref{fig:lattice}.
The vertical edge is corresponding to the (12)-(34) edge,
and horizontal one is the (42)-(31) edge in Ref.~\cite{Hasegawa_JPSJ2011}.

%------------------------------------------------------------------------------------------------------------------
\section{Results}
%-----------------------------------------------------------------------------------------------------------------

\subsection{Edge States in ZGS and the CO Phase}

%====================
\begin{figure}[b]
 \centering
 \begin{minipage}{0.45\textwidth}
 \centering 
 \includegraphics[width=2.5in]{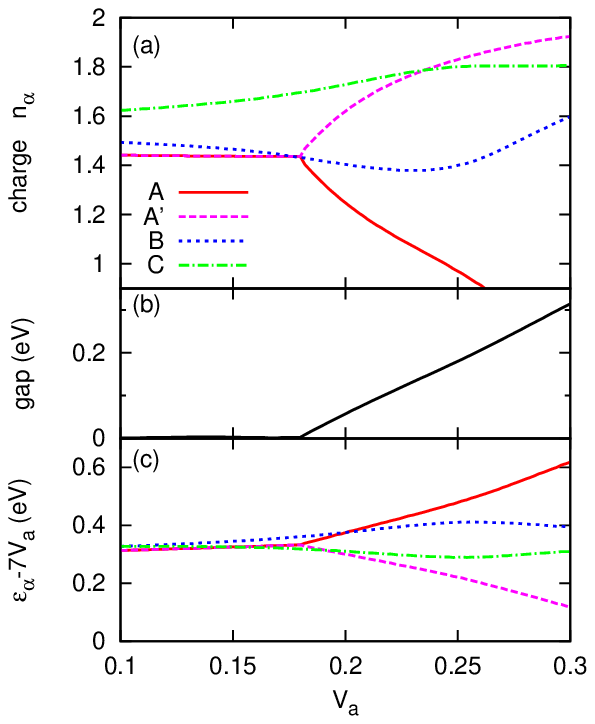}
 \caption{
 (Color online)
  The dependence on $\Va$ of
 (a) the charge density, (b) the energy gap, and (c) the effective onsite potential $\epsilon_\alpha$
  for $\Vb'=0$.
 }
 \label{fig:CO_nge}
 \end{minipage}
\hspace*{0.3in}
 \begin{minipage}{0.45\textwidth}
 \centering 
 \includegraphics[width=2.5in]{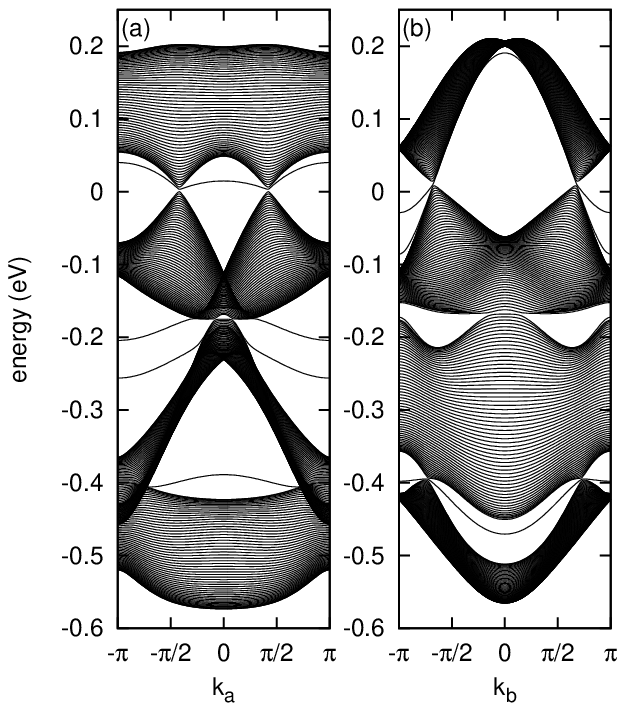}
 \caption{
  Energy spectrum of the semi-infinite system 
  with the (a) vertical  and (b) horizontal edges
  for $\Va=0.18$ eV and $\Vb'=0$.
 }
 \label{fig:CO_edge}
 \end{minipage}
\end{figure}
%====================

First, we investigate the edge state in the vicinity of the phase boundary between ZGS and CO phase.
Hereafter, we ignore the antiferromagnetic phase, 
which is not observed experimentally and almost degenerate energetically with ZGS in our model.
For small $\Va \leq 0.18$ eV and $\Vb'=0$,
charge densities on the A and A' molecules are equivalent to each other, $n_\mathrm{A}=n_\mathrm{A'}$,
and thus the system is in ZGS preserving the inversion symmetry.
As shown in Fig.~\ref{fig:CO_edge},
the energy spectrum for the semi-infinite system with the vertical or horizontal edges
has one isolated band around the Fermi energy,
which consist of the state localizing on the edges.
This result agrees with the spectrum under the artificial onsite potentials 
$\epsilon_\mathrm{A}=\epsilon_\mathrm{A'}$ and $\epsilon_\mathrm{B} \neq \epsilon_\mathrm{C}$
\cite{Hasegawa_JPSJ2011} .

%====================
\begin{figure}[t]
 \centering 
 \includegraphics[width=2in]{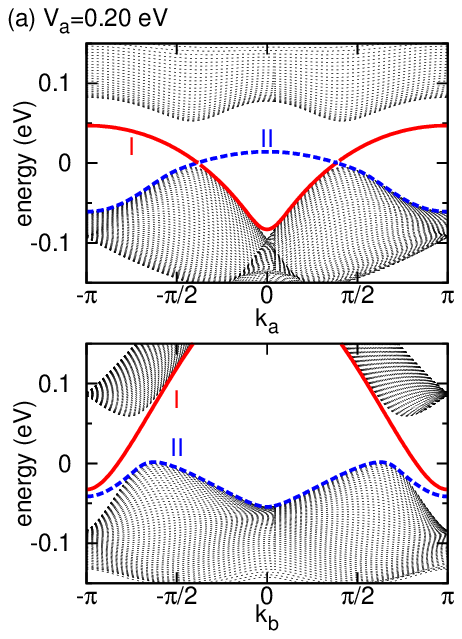}
 \includegraphics[width=2in]{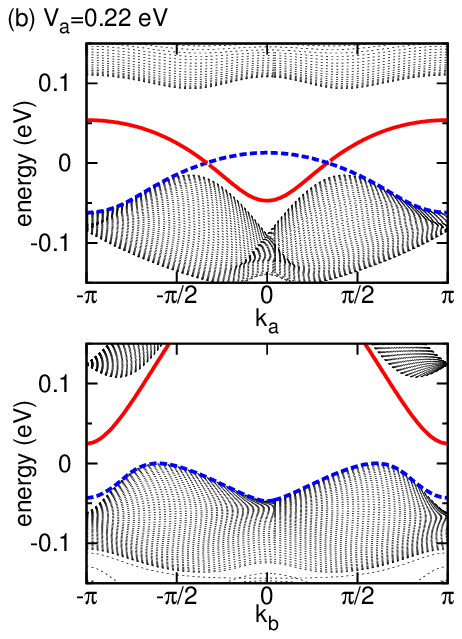}
 \includegraphics[width=2in]{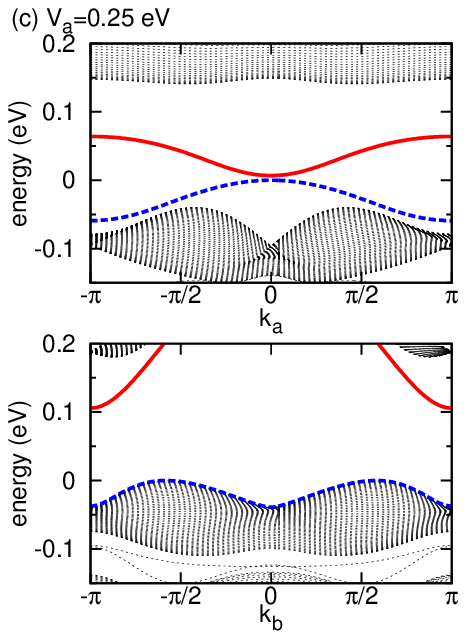}
 \caption{
 (Color online)
  Energy spectrum near the Fermi energy of the semi-infinite system for $\Vb'=0$.
 (a) $\Va=0.20$, (b) $\Va=0.22$, (c) $\Va=0.25$ eV.
 All the upper (lower) figures are for the system with the vertical (horizontal) edge.
 The red solid and blue dashed lines indicate the band I and II, respectively.
 The black dotted lines are the bulk energy bands belonging to the Bloch state.
 }
 \label{fig:CO_Cedge}
\end{figure}
%====================

%====================
\begin{figure}[b]
 \centering 
 \includegraphics[width=3in]{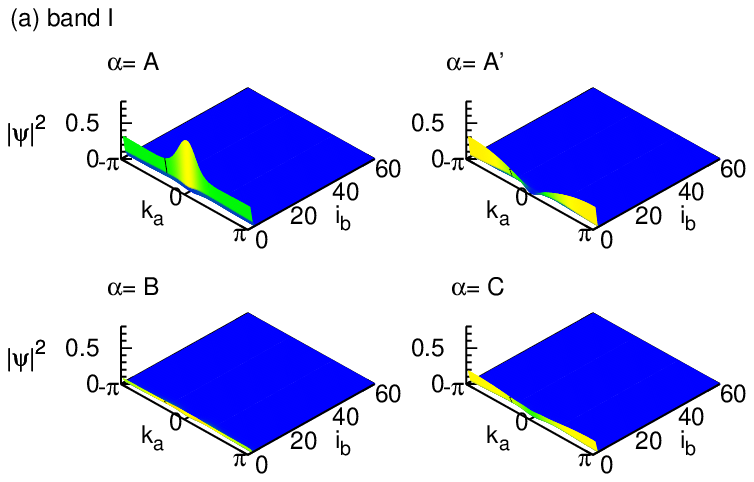}
 \includegraphics[width=3in]{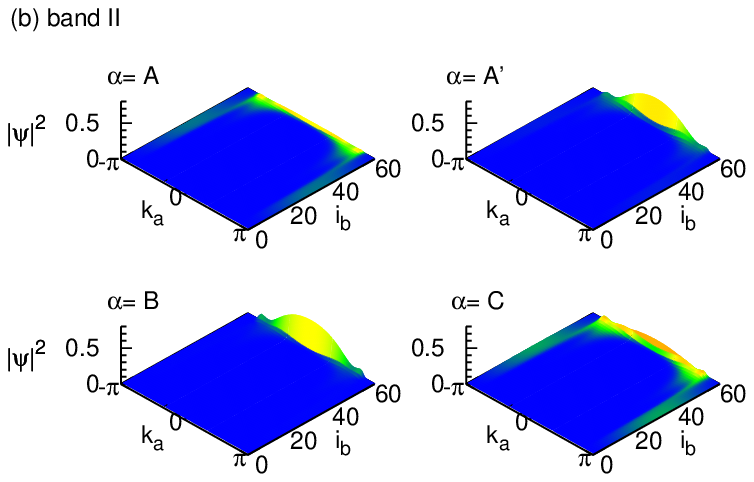}
 \caption{
 (Color online)
 The squares of the absolute values of $\psi_{k_\mathrm{b}\nu\alpha\sigma}$ 
 for the bands (a) I and (b) II for the system with the vertical edge at $\Va=0.22$ eV.
 }
 \label{fig:CO_rho}
\end{figure}
%====================

With increasing $\Va$, the phase transition occurs at $\Va=0.18$ eV.
The stripe CO phase with the order $n_\mathrm{C}>n_\mathrm{A'}>n_\mathrm{B}>n_\mathrm{A}$,
which is corresponding to the experimental observation \cite{Takahashi_SM2003},
becomes stable for $0.18<\Va<0.24$  as shown in Fig.~\ref{fig:CO_nge} (a).
Although the edge states still exist at the Fermi energy in the CO phase, 
now they consist of two bands [Fig.~\ref{fig:CO_Cedge}].
We classify them into
(i) the band I localizing on the $i_b=1$ ($i_a=N_a$) side and
(ii) band II on the $i_b=N_b$ ($i_a=1$) side for the system with the vertical (horizontal) edges,
which positions can be confirmed in the weight of the eigen vectors $|\psi_{\nu,i_b,\alpha}(k_a)|^2$
plotted in Fig.~\ref{fig:CO_rho}.
The right-side edge state, band I, has the maximum weight on the molecule A [Fig.~\ref{fig:CO_rho}(a)],
while the left-side edge II on A' and B [Fig.~\ref{fig:CO_rho}(b)].
The band I and II form the gapless band structure in the system with the vertical edges for $\Va<0.25$ eV, i.e.,
in the vicinity of the phase baoundary [Fig.~\ref{fig:CO_Cedge}(a) and (b)].
With increasing $\Va$, however,
the energy of the band I increases, and the edge-state spectrum change to be gapped for $\Va>0.25$ eV 
[Fig.~\ref{fig:CO_Cedge}(c)],
because the effective potential $\epsilon_\alpha$ becomes large on the charge-poor site A 
as shown in Fig.~\ref{fig:CO_nge} (c). 
The system with the horizontal edges becomes gapped with more smaller $\Va$.

\subsection{the Interaction-Driven Flux Phase}

As mentioned before,
when the Fock term gives the imaginary part to the effective transfer energies,
${\tilde t}^\mathrm{\ c}_{i,\alpha;j,\beta}$ or ${\bm {\tilde t}}^\mathrm{\ s}_{i,\alpha;j,\beta}$,
it plays the same role as the Peierls phases in magnetic fields
or the spin-orbit coupling in the Kane-Mele model\cite{Kane_PRL2005}.
Thus it can give rise to the interaction-driven topological insulator phase,
as studied in several lattices with Dirac electrons and inter-site Coulomb repulsions
\cite{Raghu_PRL2008,Sun_PRL2009,Liu_PRB2010,Wen_PRB2010,Kurita_JPSJ2011}.
On the \ET \ lattice, however, $\Va$ and $\Vb$
do not yield the complex Fock terms at any parameter region as far as our examination,
maybe due to the low symmetry of the lattice and the intrinsic charge disproportionation.

%====================
\begin{figure}[t]
 \centering
 \begin{minipage}{0.45\textwidth}
 \centering 
 \includegraphics[width=2.5in]{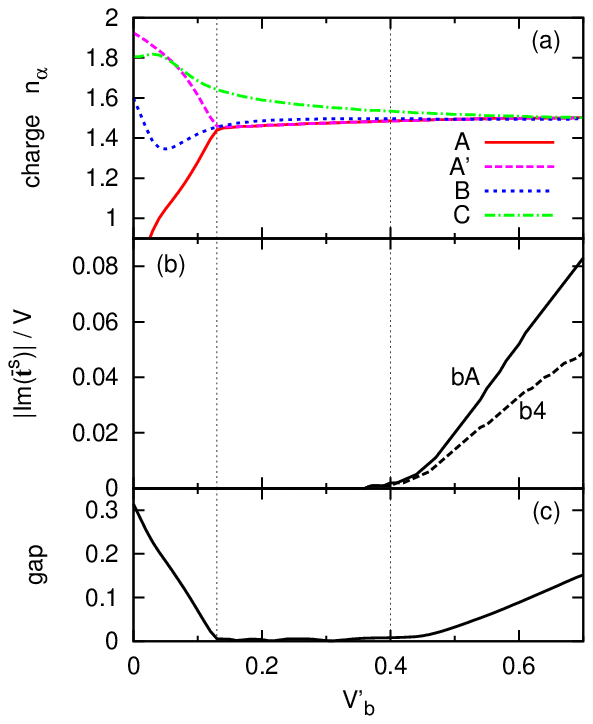}
 \caption{
 (Color online)
  (a) The charge profile,
  (b) The imaginary part of the spin-dependent effective transfer
  $| \mathrm{Im}\ ({\bm {\tilde t}}^\mathrm{\ s}_{bA})|/\Vb'$ and
  $| \mathrm{Im}\ ({\bm {\tilde t}}^\mathrm{\ s}_{b4})|/\Vb$, and
  (c) the energy gap
  for $\Va=0.30$ eV.
 }
 \label{fig:TMI_nim}
 \end{minipage}
\hspace*{0.2in}
 \begin{minipage}{0.45\textwidth}
 \centering 
 \includegraphics[width=2.5in]{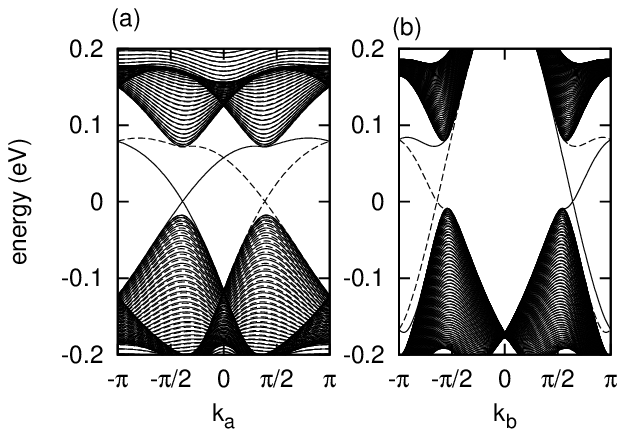}
 \includegraphics[width=2.5in]{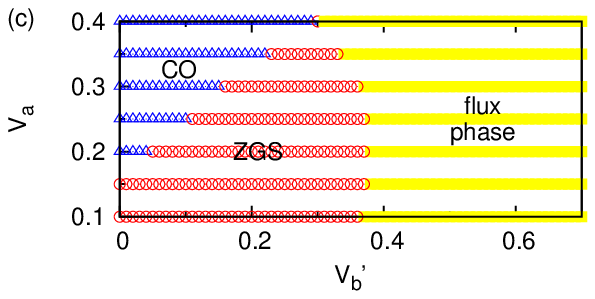}
 \caption{
 (Color online)
 The energy spectrum for (a) the vertical and (b) horizontal edges.
 The solid (dashed)lines indicate the spectral for the up (down) spin
  for $\Va=0.30$ and $\Vb'=0.60$ eV.
  (c) The phase diagram on the $\Vb'-\Va$ plane.
 }
 \label{fig:TMI_edge}
 \end{minipage}
\end{figure}
%====================

However, the large next-nearest-neighbor Coulomb repulsion $\Vb'$ can yield the complex Fock terms
as shown in Fig.~\ref{fig:TMI_nim},
where we calculate the mean fields with keeping the translational symmetry.
With increasing $\Vb'$ at $\Va=0.30$ eV,
the charge density goes to be uniform [Fig.~\ref{fig:TMI_nim}(a)],
and the imaginary part of the Fock terms appear at $\Vb'>0.40$ eV preserving the inversion symmetry.
In Fig.~\ref{fig:TMI_nim} (b)
we show the magnitude of the imaginary part of the spin-dependent effective transfer energy
between the two neighboring A molecules along the b axis,
${\bm {\tilde t}}^\mathrm{\ s}_{bA} \equiv {\bm {\tilde t}}^\mathrm{\ s}_{iA,i+1A}$,
and between the molecules A ($=$A') and C, ${\bm {\tilde t}}^\mathrm{\ s}_{b4}$,
as measured in $\Vb'$ and $\Vb$, respectively.
Other spin-dependent transfer energies also have the imaginary parts,
but all the charge sector ${\tilde t}^\mathrm{\ c}_{i\alpha,j\beta} $ remain real number.
In this parameter region $\Vb'>0.40$ eV,
the semi-infinite system has the spin-dependent edge spectrum at the Fermi energy 
as shown in Figs.~\ref{fig:TMI_edge} (a) and (b),
from which, however, it is found that the spin Chern number is even number in this phase.
Thus this phase is not topologically protected, but
one kind of flux phases composing spin currents.
Figure \ref{fig:TMI_edge} (c) is the summarized phase diagram.

\section{Summary and Discussion}

In this paper we study the edge states in \ET \ with the electron-electron interactions.
We show that the edge states exist in the CO phase and are gapless 
only in the vicinity of the phase boundary between ZGS and the CO phase.
With increasing $\Va$, it becomes gapped due to the effective onsite potential
yielded by the Hartree mean field.
These edge states are not topologically protected.
However,
according to the discussion about the robustness of the edge modes of graphene nanoribbons in the presence of 
a staggerd potential, which is helical with respect to their valley indices 
\cite{Yao_PRL2009,Qiao_PRB2011},
the edge states in the CO phase of \ET \ is also expected to be robust against scattering from smooth disorder potentials,
and may be able to contribute the transport properties.
Moreover, we show that
the strong next-nearest repulsion along the b axis yields the complex spin-independent effective transfer energies
preserving the inversion symmetry, and makes the peculiar flux phase to be stable
through the same mechanism as topological Mott insulator phase \cite{Raghu_PRL2008}.

%-----------------------------------------------------------------------------------------------------------------
%-----------------------------------------------------------------------------------------------------------------
%-----------------------------------------------------------------------------------------------------------------

\bibliographystyle{jpsj}
\bibliography{ref.bib}

% \begin{thebibliography}{9}
% \bibitem{jpsj} The abbreviation for JPSJ must be ``J. Phys. Soc. Jpn." in the reference list.
% \bibitem{instructions} More abbreviations of journal titles are listed in ``Instructions for Preparation of Manuscript", which is available at our Web site (http://jpsj.ipap.jp).
%\bibitem{format} F. Author, S. Author, and T. Author: Abbreviated journal title \textbf{volume in bold face} (year of publication) initial page or article ID.
% \end{thebibliography}

\end{document}